\documentclass[aps,prb,twocolumn,amssymb,amsfonts,groupedaddress,floatfix]{revtex4-1}
\usepackage{epsfig}
\usepackage{amsmath}
\usepackage{bm}
\usepackage{longtable}
\usepackage{times}
\usepackage{amssymb}
\usepackage{color}

\newcommand{\beq}{\begin{equation}}
\newcommand{\eeq}{\end{equation}}
\newcommand{\bea}{\begin{eqnarray}}
\newcommand{\eea}{\end{eqnarray}}
\newcommand{\ket}[1]{\left|#1\right>}
\newcommand{\bra}[1]{\left< #1 \right|}
\newcommand{\braket}[2]{\langle #1 | #2 \rangle}

\begin{document}
\title{Extended orbital modeling of spin qubits in double quantum dots}
\author{Zack White}
\author{Guy Ramon}
\email{gramon@scu.edu}
\affiliation{Department of Physics, Santa Clara University, Santa Clara, CA 95053}
\begin{abstract}

Orbital modeling of two electron spins confined in a double quantum dot is revisited. We develop an extended Hund Mulliken approach that includes excited orbitals, allowing for a triplet configuration with both electrons residing in a single dot. This model improves the reliability and applicability of the standard Hund Mulliken approximation, while remaining largely analytical, thus it enables us to identify the mechanisms behind the exchange coupling dynamics that we find. In particular, our calculations are in close agreement with exchange values that were recently measured at a high interdot bias regime, where the double occupancy triplet configuration is energetically accessible, demonstrating reduced sensitivity to bias fluctuations, while maintaining the large exchange needed for fast gating.

\end{abstract}


\maketitle

\section{Introduction}
\label{Introduction}

Spins of electrons confined in gate-defined lateral quantum dots (QDs) are a promising realization of a qubit, due to their scalability and relative isolation from their host material, as compared with the charge degree of freedom. In recent years, a remarkable progress has been made in the coherent manipulation of single-spin,\cite{Koppens_Nature06, Nowack_Science07} two-spin,\cite{Petta_Science05,Bluhm_PRL10} and three-spin\cite{Laird_PRB10,Gaudreau_Nature11} qubits.

The exchange interaction ($J$) between electron spins is a central component in all spin-based qubits. In their original proposal, Loss and DiVincenzo envisioned using gate voltage to control the exchange interaction between two electrons localized in neighboring QDs.\cite{Loss_PRA98} The high tunability of $J$ that is traditionally obtained by changing the bias ($\varepsilon$) between the two dots, provides a subnanosecond control handle. In the context of single-spin qubits, $J$ provides a fast and accessible two-qubit coupling gate. In contrast, single-spin rotations require coupling to the small magnetic moment of the electron, which is much more challenging. Combining on-chip micromagnets that create field gradients across each QD,\cite{Pioro_Nature08} with electrical control over the exchange interaction,\cite{Petta_Science05} Tarucha's group demonstrated a universal set of gates on solid state single-spin qubits, albeit with single-spin rotation times are still 2-3 orders of magnitude longer than exchange gates.\cite{Brunner_PRL11}

The challenging manipulation of single spin qubits has prompted a number of proposals to encode the logical qubit states into two-spin singlet ($S$) and unpolarized triplet ($T_0$) states.\cite{Lidar_PRL98,Levy_PRL02,Taylor_Nature05} In these devices, Pauli spin blockade is used to initialize the qubit in a  doubly occupied singlet state, $J$ provides single-qubit rotations about the $z$ axis, and inhomogeneous nuclear spin polarization generates a magnetic field gradient that provides $x$-axis rotations.\cite{Foletti_Nature09} Two-qubit gates between $S-T_0$ qubits in neighboring two double QDs were proven to be more challenging but have also been demonstrated.\cite{Weperen_PRL11,Shulman_Science12}

While encoding the qubit states in exchange-coupled electron spins alleviates the challenging tasks of single-spin addressability and control, it renders the $S-T_0$ qubit vulnerable to decoherence from charge noise, since $J$ is electrostatic in nature and its coupling to the fluctuating charge environment (e.g., through its interdot bias dependence) is much stronger than the spin-orbit-mediated charge coupling of single-spin qubits.\cite{Hu_PRL06,Culcer_APL09,Ramon_PRB10}
The sensitivity of $J$ to bias fluctuations is heightened with increased bias, due to the very different charge distributions of the triplet and hybridized singlet in this regime. This results in increased susceptibility of the qubit to charge noise and has been a long lasting problem of these QD devices, as the desired fast gating achieved in the positive bias regime comes at the price of increased decoherence rates. Indeed, the realization of a controlled-PHASE gate between two $S-T_0$ qubits was made possible by identifying charge noise as the main obstacle and mitigating it using a spin echo pulse along the $x$ axis.\cite{Shulman_Science12}

In a set of experiments targeted at characterizing both low- and high-frequency components of charge noise, Dial {\it et al.} first operated an $S-T_0$ qubit at a high bias regime, where excited orbitals and thus a doubly occupied triplet configuration become energetically accessible.\cite{Dial_PRL13} In this regime, both singlet and triplet states are hybridized, and their charge distributions are much more similar, giving rise to reduced $dJ/d\varepsilon$ while maintaining large $J$. As expected,  reduced sensitivity to charge noise was manifested at this new regime by enhanced quality of coherent exchange oscillations.

The main goal of the current work is to develop an orbital model that captures the electronic states at this high-bias operating regime. Rather than employing all-numerical approaches like Configuration-Interaction (CI)\cite{Nielsen_PRB10} or exact diagonalization,\cite{Hiltunen_PRB14} we develop an extended Hund-Mulliken (HM) approach that allows us to derive analytical results, while rendering the important orbital features of the system correct within a useful range of parameters, as specified bellow. In addition to using our model to study the high-bias regime, we explore the dependence of $J$ on magnetic field and double dot geometry. Our analytical treatment points at the main mechanisms behind the exchange behavior and allows us to identify useful working positions that may improve electrical control of QD spin qubits.

\section{Extended orbital model for singlet-triplet qubits in a double QD }
\label{sec:model}

The physical system we consider consists of two electrons localized in a pair of laterally-coupled QDs, whose singlet and triplet spin configurations serve as the qubit computational basis states. For concreteness we employ parameters relevant for GaAs quantum dots, but our approach can be applied directly to other semiconductor QD materials, and in particular to Si\cite{Tyryshkin_Nature12} or ${\rm Si}/{\rm SiGe}$,\cite{Maune_Nature12,Kim_Nature14} as long as a single-valley calculation is justified, i.e., when the considered device has a sufficiently large valley splitting and a uniform ground valley state composition, where intervalley matrix elements vanish.\cite{Culcer_APL09,Li_PRB10} In such a case, the Hamiltonian we consider below effectively does not couple the ground valley state to higher energy valley states and our model is directly applicable for Si-based QDs. We note, however, that the larger effective mass and dielectric constant in Si, generate larger Coulomb couplings and smaller kinetic energy, as compared with GaAs, resulting in more stringent conditions for the validity of the HM model.

\subsection{System Hamiltonian}
\label{sec:H}

Taking the magnetic ($B$) and electric ($E$) fields along the $z$- and $x$-axis, respectively, the Hamiltonian describing the coupled QDs includes single-particle, two-particle (Coulomb) and Zeeman terms and is given by
\begin{eqnarray}
H \! &\! =\! &\! H_{\rm orb}+H_Z=\sum \limits_{i=1}^{2} H_i^{\rm SP}+H_C+H_Z, \label{H} \\
H_i^{\rm SP}\!&\!=\!&\! \frac{1}{2m}\left[\mathbf{p_i}-\frac{e}{c}\mathbf{A}(\mathbf{r_i})\right]^2+e x_i E + V(\mathbf{r_i}),\label{HSP} \\
H_C\!&\!=\!&\! \frac{e^2}{\kappa |\mathbf{r_1}-\mathbf{r_2}|}, \hspace{0.5 cm} H_Z=\frac{g \mu_B}{\hbar} \mathbf{B} \cdot \sum_i \mathbf{S_i}.
\end{eqnarray}
The single-particle Hamiltonian, $H_i^{\rm SP}$, describes the dynamics of a single electron confined in the $x-y$ plane, under applied fields, where $B$ is coupled to the electron charge through a vector potential ${\bf A}({\bf r})=\frac{1}{2}B(-y,x,0)$. Taking GaAs parameters, the electron's effective mass as $m$=0.067$m_e$, and the dielectric constant and effective g-factor are $\kappa=13.1$ and $g =-0.44$, respectively.

\begin{figure}[t]
\epsfxsize=0.75\columnwidth
\vspace*{0.0 cm}
\centerline{\epsffile{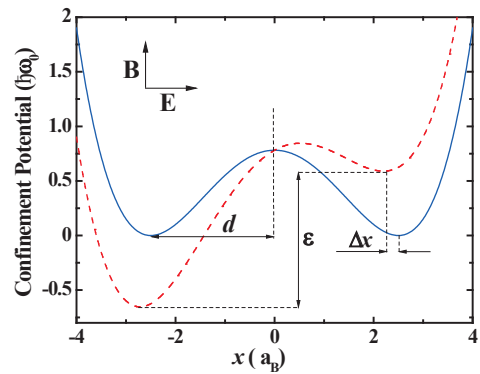}}
\vspace*{-0.2 cm}
\caption{(Color online) A cut along the $y=0$ plane of the quartic confinement potential for $d=2.5$, with (dashed-red line) and without (solid-blue line) interdot bias. In the limit where $\Delta x \ll d$, the bias term approximately shifts the two potential minima by $-\Delta x$, and induces an approximate energy difference $\varepsilon$, both defined in the main text. The exact bias-shifted single-particle energies that are used in our calculations are given in appendix \ref{sec:app_Helements}.}
\label{Fig1}
\end{figure}

We consider a two-dimensional quartic confinement potential\cite{Burkard_PRB99} to model the two nominally identical QDs:
\begin{equation}
V(x,y)=\frac{\hbar \omega_0}{2} \left[\frac{1}{4d^2}(x^2-d^2)^2+y^2\right], \label{quartic}
\end{equation}
where $d$ is the dimensionless half interdot distance, measured in effective Bohr radius of a harmonic dot, $a_B=\sqrt{\hbar/m \omega_0}$, and the coordinates henceforth are measured in $a_B$ units. The gate-voltage-induced electric field, $E$, in Eq.~(\ref{HSP}) is directed along the line connecting the centers of the two dots and splits the single particle energies through tilting of the confinement potential, as well as shifting the orbitals. Figure \ref{Fig1} qualitatively depicts the effects of interdot bias on the confinement potential. For all experimentally relevant parameters, we can assume $\Delta x \equiv eE a_B/\hbar \omega_0 \ll d$, corresponding to electric fields $E \ll 0.8$ MV/m for $\hbar \omega_0=5$ meV and $d=2.5$. In this limit, the position shift and energy difference of the well minima are approximately given by the quantities $\Delta x$, and $\varepsilon \equiv 2\hbar \omega_0 d \Delta x$, respectively (see Fig.~\ref{Fig1}).

$H_C$ represents the bare Coulomb interaction between the two electrons, and we neglect screening effects that are expected to be minimal in these few-electron QD devices. We note that the Zeeman splitting is much smaller than the orbital energies we consider. For $\hbar \omega_0=5$ meV --- the nominal confinement we choose unless otherwise specified --- $g \mu_B B/ \hbar \omega_0 \lesssim 0.03$ ($\mu_B$ is the Bohr magneton), for magnetic fields up to 6T, well beyond typical applied fields. This allows us to ignore the Zeeman term in the orbital energy calculations, adding it only in the effective spin Hamiltonian. Finally, spin-orbit coupling for confined electrons is several orders of magnitude smaller than the orbital energy scale, and can be safely ignored in this context.

The $2D$ quartic potential in Eq.~(\ref{quartic}) assumes infinite confinement in the growth ($z$) direction, appropriate for typical gate-defined QD structures, and enables us to approximate the Coulomb interactions using 2-D integrals. It has been demonstrated experimentally that single-dot spectra in GaAs are adequately described by a parabolic potential with $\hbar \omega_0$ on the order of a few meV.\cite{Tarucha_PRL96} Our main focus here is on weakly-coupled QDs with relatively large separation. At the limit of $d \gg 1$ the quartic potential separates into two harmonic wells centered around $\pm d$, which provides a convenient starting point for orbital matrix elements calculations. We stress that confinement in real QD devices may be substantially different from our simple model. A detailed knowledge of the device electrostatic structure can enable accurate orbital modeling, using e.g., numerical Schr\"{o}dinger-Poisson CI methods (see for example Ref.~\onlinecite{Stopa_NL08}, where the CI basis was derived from device-specific density functional theory calculations). Instead, our approach here is to use a simple confinement potential that allows us to obtain analytical results, elucidating the important physical mechanisms behind orbital dynamics in these devices.

Other confinement potentials have been used to model double QDs, including a linear combination of three Gaussians (accounting for the two wells and central barrier),\cite{Hu_PRA00} a biquadratic potential,\cite{Pedersen_PRB07,Nielsen_PRB10,Li_PRB10,Barnes_PRB11} and a more realistic finite quadratic potential, used in conjunction with matched variational orbitals in a Heitler-London calculation.\cite{Saraiva_PRB07, Li_PRB10} The simpler quartic and biquadratic potentials both feature unrealistic infinite confinement as $r \rightarrow \infty$. In addition, the biquadratic potential underestimates the interdot overlap due to its unphysical kink in the central barrier, whereas the quartic potential has different confinement energies in the two dots under biased configuration. At sufficiently strong bias, the quartic potential results in a single well, thus it was argued that the biquadratic potential is better suited to model biased double dots.\cite{Li_PRB10} Motivated to obtain analytical results, we nevertheless adopt the quartic potential, contending that even in the large-bias regime considered below, we are close, but still below the single-well threshold.

Another caveat shared by the quartic and biquadratic potentials is that the barrier hight governing the tunnel coupling between the dots is determined by the interdot distance, whereas in experiments the tunnel barrier is controlled by a gate voltage separately from the interdot distance.\cite{Waugh_PRL95} Whereas most experiments implement exchange control using bias detuning, and are thus reasonably modeled by our biased quartic potential, recent experiments in GaAs\cite{Martins_PRL16} and Si\cite{Reed_PRL16} QDs have successfully implemented symmetric exchange control by independently tuning the central barrier height, resulting in improved immunity to charge noise. This additional control can be modeled by refining our confinement potential to include a Gaussian term that provides a separate handle for the interdot potential barrier, in the spirit of Ref.~\onlinecite{Hu_PRA00}, but is outside the scope of the current work.

\subsection{Extended Hund Mulliken approach}
\label{sec:HM}

Casting the orbital Hamiltonian as an effective spin Hamiltonian, $J\mathbf{S_1} \cdot \mathbf{S_2}$, the exchange energy, $J$, is found by diagonalizing the singlet and triplet subspaces of the Hamiltonian and taking the difference between the lowest-lying singlet and triplet states, $J=E_t -E_s$. There is an infinite number of single-dot orbitals from which the two-electron states are built, thus all calculational approaches, including full CI, inevitably use a truncated basis.

The Heitler-London approximation is the simplest approach, in which only the ground-state orbitals are used to build the two separated symmetric and antisymmetric two-electron states.\cite{Mattis_Springer} Singlet-triplet qubits operate at biased configuration and even at moderate bias, the Heitler-London approximation breaks down, as double occupancy states become energetically favorable. The similarity of electronic states in a gate-defined double dot to molecular orbitals suggests the Hund-Mulliken (HM) approximation as an appropriate approach.\cite{Mattis_Springer,Burkard_PRB99} In this approximation the single-particle basis states still comprise of only the $s$ orbital in each dot, but doubly-occupied two-electron states are allowed, resulting in a four dimensional orbital Hilbert space with three singlets: $S(2,0)$, $S(0,2)$, $S(1,1)$, and one triplet: $T(1,1)$ (due to Pauli exclusion). Here the two numbers denote the number of electrons occupying the left and right dot. Considering positive interdot bias, electron tunneling is induced from the right to left dot. The hybridization of the $S(1,1)$ and $S(2,0)$ singlets dramatically lowers the ground singlet state and in the absence of such triplet hybridization, the resulting exchange increases by several orders of magnitude (negative bias will similarly induce $S(1,1)$ - $S(0,2)$ hybridization with the same resulting exchange).

As discussed in the introduction, we are interested in extending the standard HM model to capture the orbital dynamics in the high-bias regime, where $p$ orbitals and thus doubly occupied triplet states become energetically accessible. Rather than employing an $sp$-hybridized Heitler-London basis\cite{Burkard_PRB99} (that still excludes double occupancy states and is thus unsuitable for this biased regime), or performing an all-numerical CI calculation with extended state basis, we opt to keep the Hilbert space minimal by including only the lowest excited orbital in each dot, while still keeping double occupancy states. This choice limits the validity regime of our model, but as shown below, we are still able to qualitatively capture the exchange dynamics at biases well beyond the triplet anticrossing,\cite{Dial_PRL13} covering a wide range of working positions employed in current experiments with QD devices. The validity and reliability of our exchange calculations are discussed in appendix \ref{sec:app_truncate}. We note that strictly speaking an HM calculation includes only $s$ orbitals, but to avoid confusion with the more commonly used $sp$-hybridized Heitler-London approximation, we refer to our approach as \emph{extended Hund-Mulliken}.

A standard single-particle state basis, from which the approximate two-particle state solutions for the orbital Hamiltonian in Eq.~(\ref{H}) can be constructed, are the Fock-Darwin solutions to the two-dimensional parabolic potential (for a derivation see, e.g., Ref.~\onlinecite{Barnes_PRB11}). The ground and first excited (so called $p_-$) orbitals under magnetic field, centered at the minima of each (unbiased) well, $\pm d$, are given (in $a_B^{-1}$ units) by:
\begin{eqnarray}
\varphi_{\pm}^g\! & \!=\! &\! \sqrt{\frac{b}{\pi}}e^{-\frac{b}{2} \left[( x\mp d)^2+y^2\right]} e^{\mp iyd \sqrt{b^2-1}} \notag \\
\varphi_{\pm}^e\! & \!=\! &\! \frac{b}{\sqrt{\pi}}e^{-\frac{b}{2} \left[( x\mp d)^2+y^2\right]} e^{\mp iyd \sqrt{b^2-1}} \left( x \mp d - iy \right),
\label{FD}
\end{eqnarray}
and their energies are $\hbar \omega$ and $2\hbar \omega -\hbar  \omega_L$, respectively. Here, $\omega=b\omega_0$, where $b=\sqrt{1+\omega_L^2/\omega_0^2}$ is the magnetic compression factor, and $\omega_L=eB/(2mc)$ is the Larmor frequency. Henceforth, we state energies in $\hbar \omega_0$ units, so that the ground and first excited energies read $b$, and $2b-\sqrt{b^2-1}$, respectively. In the presence of an electric field, the above orbitals are both shifted by $\Delta x$.

\begin{figure}[t]
\epsfxsize=0.75\columnwidth
\vspace*{0.0 cm}
\centerline{\epsffile{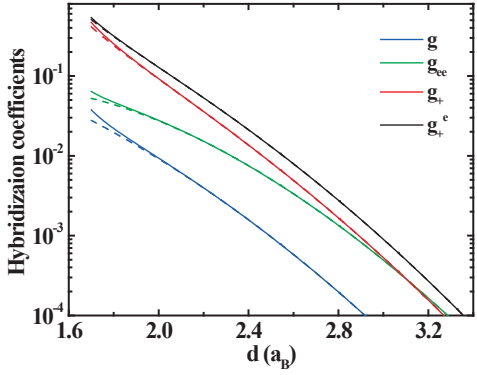}}
\vspace*{0.0 cm}
\caption{(Color online) Orbital hybridization coefficients vs.~half interdot separation $d$. Solid lines depict the exact numerical solution to Eqs.~(\ref{geq}) and dashed lines show the approximate solution given by Eqs.~(\ref{gsol}).}
\label{Fig2}
\end{figure}

While the Fock-Darwin solutions are orthogonal within each dot, interdot wavefunctions have the following overlaps:
\begin{eqnarray}
S&\equiv& \braket{\varphi_+}{\varphi_-}=e^{-d^2\left(2b-\frac{1}{b}\right)} \notag \\
S_{\pm} & \equiv& \braket{\varphi_{\pm}}{\varphi_{\mp}^e}=\pm \left(b+\sqrt{b^2-1}\right) S \notag \\
S_{ee}& \equiv & \langle \varphi_+^e | \varphi_-^e \rangle=\left[1-\frac{d^2}{b}\left(b+\sqrt{b^2-1}\right)^2\right]S.
\label{S}
\end{eqnarray}
To facilitate the construction of the two-electron Hamiltonian matrix elements, we first orthonormalize the four basis states in Eqs.~(\ref{FD}). The orthonormalized states are given by
\begin{eqnarray}
\Phi_{\pm}^g \!&\!=\!&\! \frac{1}{N_g}\left(\varphi_{\pm}^g -g \varphi_{\mp}^g - g_{\pm} \varphi_{\mp}^e \right) \notag \\
\Phi_{\pm}^e\!&\!=\!&\! \frac{1}{N_e}\left(\varphi_{\pm}^e - g_{ee} \varphi_{\mp}^e -g_{\pm}^e \varphi_{\mp}^g  \right).
\label{Phi}
\end{eqnarray}
Each orthonormalized state is comprised primarily of one Fock-Darwin orbital, with additional contributions corresponding to the overlap between the primary orbital and the two orbitals in the other dot. The normalization constants are found to be $N_g=(1+g^2+g_{\pm}^2-2g_{\pm}S_{\pm}-2gS)^{-1/2}$ and $N_e=(1+g_{ee}^2+g_{\pm}^{e2}+2g_{\pm}^e S_{\pm}-2g_{ee}S_{ee})^{-1/2}$. The hybridization coefficients are calculated from the orthogonalization conditions of the wavefunctions, Eqs.~(\ref{Phi}):
\begin{eqnarray}
\braket{\Phi_{\pm}}{\Phi_{\mp}} \!&\!=\!&\! S-2g+g^2S+g(g_{\pm}S_{\pm}+g_{\mp}S_{\mp})+ \notag \\ && g_{+}g_{-}S_{ee} = 0 \notag \\
\braket{\Phi_{\pm}^e}{\Phi_{\mp}^e} \!&\!=\!&\! S_{ee}-2g_{ee}+ g_{ee}^2S_{ee} +g_{ee}(g_{\mp}^e S_{\pm}+g_{\pm}^e S_{\mp})+  \notag \\ && g_{+}^eg_{-}^eS = 0 \notag \\
\braket{\Phi_{\pm}}{\Phi_{\mp}^e} \!&\!=\!&\! S_{\pm}-g_{\pm}-g_{\pm}^e\!+g_{\mp}^e(gS+g_{\pm}S_{\pm}\! )+gg_{ee}S_{\mp}+ \notag \\ && g_{\pm}g_{ee}S_{ee} = 0 \notag \\
\braket{\Phi_{\pm}}{\Phi_{\pm}^e} \!&\!=\!&\! -g_{\pm}^e S -g_{ee}S_{\pm} -gS_{\mp}+gg_{\pm}^e-g_{\pm}S_{ee}+ \notag \\ && g_{\pm}g_{ee} = 0.
\label{geq}
\end{eqnarray}
Symmetry considerations determine that $g_-=-g_+$ and $g_-^e=-g_+^e$. These coupled nonlinear equations cannot be solved analytically, but for sufficiently large interdot distance, where overlaps are small, the $g$ coefficients are well approximated by their first order solution:
\begin{eqnarray}
g &\approx &  \frac{S}{2}, \hspace{2.1 cm}  g_{ee} \approx \frac{S_{ee}}{2} \notag \\
g_{\pm} &\approx & S_{\pm} \frac{2S - S_{ee}}{S+S_{ee}}, \hspace{5 mm}  g_{\pm}^e \approx S_{\mp} \frac{2S_{ee} - S}{S+S_{ee}}.
\label{gsol}
\end{eqnarray}
A comparison between these approximate solutions and the numerically obtained exact values is shown in Fig.~\ref{Fig2} as a function of half interdot separation $d$. We observe that the approximate solutions are accurate down to $d\gtrsim 2$, covering device geometries employed in experiments with GaAs gate-defined QDs. For $d \lesssim 2$ and typical confinement energies, the HM model breaks down anyways (see appendix \ref{sec:app_truncate}), thus Equations (\ref{gsol}) can be safely used within our model's validity regime.

There are ten singlets and six triplets that can be constructed from the four orthogonalized states in Eqs.~(\ref{Phi}). We reduce the Hilbert space by eliminating two-particle states comprising two excited orbitals, which are energetically inaccessible to the same extent as states comprising a ground orbital and the second excited ($p_+$) orbital, originally excluded from our model. The resulting Hilbert space has a total of 12 states with seven singlets and five triplets, listed in Table \ref{Tstates}. When considering a limited bias regime (e.g., around the singlet or triplet avoided crossings), one can further reduce the Hilbert space and obtain closed-form analytical results, as we show in appendix \ref{sec:app_approx}. The matrix elements of the orbital Hamiltonian in Eq.~(\ref{H}), in the basis of the 12 states given in Table \ref{Tstates}, include single-particle energy and tunneling terms and two-particle Coulomb coupling terms. They are constructed from bare (non-orthogonalized) matrix elements whose explicit closed-form expressions are given in appendices B1, and B2, respectively. Appendix B3 details the process of obtaining the final matrix elements from these building blocks. The Hamiltonian matrix is then diagonalized numerically and the orbital eigenenergies are determined.

\begingroup
\squeezetable
\begin{table}[th]
\caption{The 12 two-particle states comprising the system's truncated Hilbert space. The three singlets and one triplet listed in the top portion of the table are the ground orbitals included in the standard HM model.\cite{Burkard_PRB99}}
\label{Tstates}
\begin{tabular}{l|l}\hline \hline
Singlets & Triplets \\
$S(0,2)\!=\!\Phi_+ \Phi_+$ & --- \\
$S(2,0)\!=\!\Phi_- \Phi_-$ & --- \\
$S(1,1)=\frac{1}{\sqrt{2}}(\Phi_- \Phi_+ \!+\! \Phi_+ \Phi_-)$ & $T(1,1)\!=\!\frac{1}{\sqrt{2}}(\Phi_- \Phi_+ \!-\! \Phi_+ \Phi_-)$ \\ \hline
$S_e(0,2)\!=\!\frac{1}{\sqrt{2}}(\Phi_+ \Phi_+^e \!+\! \Phi_+^e \Phi_+)$ & $T_e(0,2)\!=\!\frac{1}{\sqrt{2}}(\Phi_+ \Phi_+^e \!-\! \Phi_+^e \Phi_+)$\\
$S_e(2,0)\!=\!\frac{1}{\sqrt{2}}(\Phi_- \Phi_-^e \!+\! \Phi_-^e \Phi_-)$ & $T_e(2,0)\!=\!\frac{1}{\sqrt{2}}(\Phi_- \Phi_-^e \!-\! \Phi_-^e \Phi_-)$\\
$S_{ge}(1,1)\!=\!\frac{1}{\sqrt{2}}(\Phi_- \Phi_+^e \!+\! \Phi_+^e \Phi_-)$ & $T_{ge}(1,1)\!=\!\frac{1}{\sqrt{2}}(\Phi_- \Phi_+^e \!-\! \Phi_+^e \Phi_-)$ \\
$S_{eg}(1,1)\!=\! \frac{1}{\sqrt{2}}(\Phi_-^e \Phi_+ \!+\! \Phi_+ \Phi_-^e)$ & $T_{eg}(1,1)\!=\!\frac{1}{\sqrt{2}}(\Phi_-^e \Phi_+ \!-\! \Phi_+ \Phi_-^e)$ \\
\hline \hline
\end{tabular}
\end{table}
\endgroup

\section{Exchange splitting results}

We now present results of our exchange calculations and its dependence on various parameters. We note that across the range of approximation approaches and model confinement potentials, exchange values can differ by an order of magnitude,\cite{Li_PRB10} thus our results should be considered as qualitative or semi-quantitative, at best. Nevertheless, since the bias dependence of $J$ spans several orders of magnitude, we believe that the main features are correctly captured.

\subsection{High-bias regime}

\begin{figure}[t]
\epsfxsize=0.85\columnwidth
\vspace*{-0.05 cm}
\centerline{\epsffile{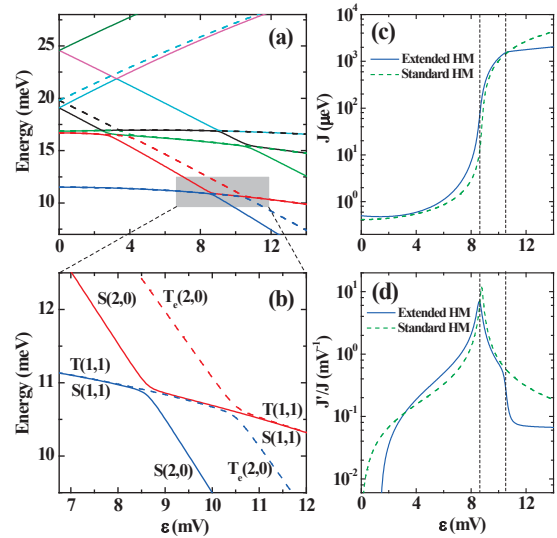}}
\vspace*{-0.1 cm}
\caption{(Color online) (a) Orbital energies of the 12 two-electron states included in our extended HM model vs.~interdot bias. Solid (dashed) lines depict singlet (triplet) energies. (b) Zoom on the lowest-lying singlet and triplet energies, exhibiting avoided crossings. The labels state the predominant content of each eigenfunction. (c) Exchange energy calculated with the current extended HM approach (solid blue line) and the original HM model of Ref.~\onlinecite{Burkard_PRB99} (dashed green line). Vertical dotted lines mark the locations of the two anticrossings, determined by equal $S(1,1)$ and $S(2,0)$ ($T(1,1)$ and $T_e(2,0)$) content in the lowest lying singlet (triplet) eigenvectors. (d) $(\partial J/\partial \varepsilon) /J$ calculated with the extended (solid blue line) and original (dashed green line) HM models. For all plots, $B=0.1$ T and the quartic confinement potential parameters are $\omega_0=5$ meV and $d=2.5$.}
\label{Fig3}
\end{figure}

An example of the energy diagram as a function of interdot bias, $\varepsilon$, is shown in figure \ref{Fig3}a, where solid (dashed) lines depict singlet (triplet) energies. Here and throughout the paper $\varepsilon$ corresponds to the energy difference between the two wells, as depicted in figure \ref{Fig1}, rather than the detuning from the $S(2,0)-S(1,1)$ degeneracy point, commonly used in many experimental works. Focusing on the lowest lying singlets (primarily comprising of $S(1,1)$ and $S(2,0)$), and triplets (primarily comprising of $T(1,1)$ and $T_e(2,0)$), figure \ref{Fig3}b depicts the expected singlet anticrossing, followed by a triplet anticrossing at a higher interdot bias. At this large positive bias, the resulting exchange energy (blue solid line in figure \ref{Fig3}c) presents marked flattening as compared with the standard HM (green dashed line), where no triplet hybridization is allowed. The high-bias $J$ flattening observed in our extended HM calculation is a direct result of the double-occupancy states included in our model for both singlet \emph{and} triplet configurations, as their charge distributions and thus their bias dependence is much more similar.

In this positive bias regime, the qubit dephasing time, $T_2^*$, has been experimentally found to be inversely proportional to $\partial J/\partial \varepsilon$, suggesting that nuclear noise is negligible and the main noise source is low-frequency voltage fluctuations.\cite{Dial_PRL13} The observed number of exchange oscillations, used as a figure of merit for their quality, is therefore proportional to $J(\partial J/\partial \varepsilon)^{-1}$ and is experimentally found to be approximately constant at the positive bias regime (consistent with the observed exponential dependence of $J$ on $\varepsilon$). The plot of $(\partial J/\partial \varepsilon)/J$ in figure \ref{Fig3}d shows a pronounced maximum at the singlet anticrossing and an abrupt drop at the triplet anticrossing to a value of $~0.07$ -- five-fold smaller than it's calculated value without a doubly-occupant triplet (dashed-green line). The experimental value extracted at this high-bias regime is $~0.06$ -- remarkably close to our calculated value, demonstrating the simultaneous existence of large $J$ and reduced sensitivity to bias fluctuations.\cite{Dial_PRL13} Overall, we find a reduction of two orders of magnitude in $(\partial J/\partial \varepsilon)/J$ from its maximal value. As was noted in Ref.~\onlinecite{Dial_PRL13}, close to the singlet anticrossing, where $\partial J/\partial \varepsilon$ is very large, $T_2^*$ is too short to be correctly captured, and the extracted values of $J$ and  plot $\partial J \partial /\varepsilon J$ are unreliable, precluding a direct comparison of our calculations with measured values.

\subsection{Exchange dependence on magnetic field and interdot bias}

\begin{figure}[t]
\epsfxsize=0.8\columnwidth
\vspace*{-0.05 cm}
\centerline{\epsffile{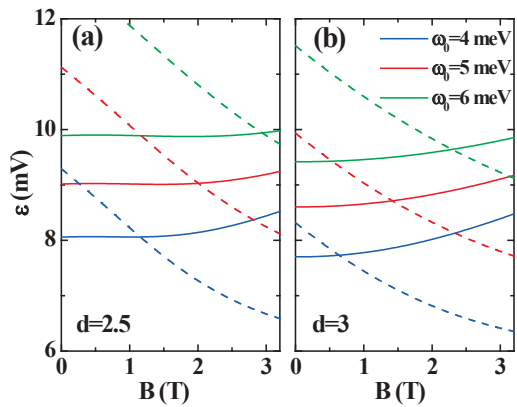}}
\vspace*{-0.1 cm}
\caption{(Color online) Bias locations of singlet (solid lines) and triplet (dashed lines) anticrossings vs.~magnetic field for several quartic confinement energies and half interdot separation of: (a) $d=2.5$, (b) $d=3$.}
\label{Fig4}
\end{figure}

The orbital energy structure shown in figure \ref{Fig3} can change significantly with different system parameters. We first study the relative bias locations of the lowest lying singlet and triplet anticrossings, as they are directly responsible for the $J$ tunability and its sensitivity to bias fluctuations. Figure \ref{Fig4} shows anticrossing bias locations vs.~$B$ for several confinement energies and two interdot distances. Generally, these anticrossings occur when electron tunneling from the right to left dot becomes energetically favorable. In a triplet configuration, the electron needs to tunnel to the excited orbital in the left dot, whereas in a singlet configuration it can tunnel to the ground orbital. One would then naively estimate the bias difference between singlet and triplet anticrossings to amount to the energy gap between the $s$ and $p_-$ orbitals, which at zero magnetic field is simply $\hbar \omega_0$. This gap is partially offset, mostly due to the reduced on-site Coulomb repulsion in the doubly-occupied triplet state ($U$ terms in table \ref{THterms}), resulting in a $\sim 1.25$ meV energy difference for $\hbar \omega_0=4$ meV, as seen in Fig.~\ref{Fig4}a.

With increased magnetic field, the ground orbital energy is raised, while the excited orbital energy is reduced. The singlet anticrossing is then mostly impacted by the increased on-site Coulomb repulsion (due to the magnetic orbital compression), resulting in a mild increase in its anticrossing bias, whereas the triplet anticrossing bias is more drastically reduced due to the reduced Fock-Darwin excited orbital energy. The order at which the singlets and triplets anticross is thus reversed at $B=1.15$ T for $\hbar \omega_0=4$ meV and at larger $B$ for stronger confinement, due to the scaling of the energy gap between the ground and excited orbitals. Increasing the interdot separation from $d=2.5$ to $d=3$ (Fig.~\ref{Fig4}b) has almost no effect on the on-site Coulomb terms at the weak-coupling regime we consider, but it reduces the interdot direct and exchange Coulomb terms by $\sim 0.2$ meV and should have therefore increased the anticrossing biases. Instead, we observe that both anticrossings occur at lower biases, and find it to be due to the $d$-dependence of the quartic-potential-related $Q$ terms in the single-particle energies, with larger effect on the excited orbital and thus on the triplet energies (see table \ref{Tsingle}). As a result, the anticrossing order reversal occurs at lower magnetic fields ($B=0.63$ T for $\hbar \omega_0=4$ meV), as seen in Fig.~\ref{Fig4}b. Since the $d$ dependence of the anticrossing locations is specific to the quartic potential details, one should not take it too seriously.

\begin{figure}[t]
\epsfxsize=0.6\columnwidth
\vspace*{0.2 cm}
\centerline{\epsffile{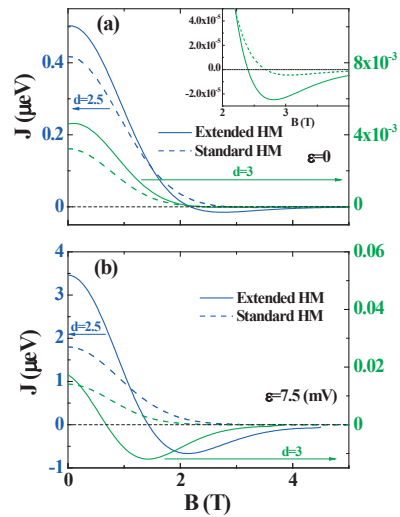}}
\vspace*{-0.2 cm}
\caption{(Color online) $J$ vs.~magnetic field calculated using our extended HM model (solid lines) and standard HM (dashed lines) at: (a) zero bias; (b) $\varepsilon =7.5$ mV. Blue lines and left axes  depict $J$ for half interdot separation of $d=2.5$, while green lines and right axes depict $J$ for $d=3$. The inset in panel (a) provides a zoom around the $S-T$ crossing for the $d=3$ case. $\omega_0=5$ meV in all plots.}
\label{Fig5}
\end{figure}

The dependence of $J$ on the magnetic field exhibits several interesting features. At larger magnetic fields, $J$ demonstrates an exponential decay due to the increased magnetic compression of the orbitals (and hence their reduced overlap), in qualitative agreement with the standard HM model. We have verified that $J>0$ at $B=0$ throughout the parameter range considered, satisfying the Lieb-Mattis theorem for a two-particle system under a symmetric potential\cite{Mattis_Springer} (see also appendix \ref{sec:app_truncate}). For a symmetric double dot ($\varepsilon=0$), figure \ref{Fig5}a shows a transition from antiferromagnetic ($J>0$) to ferromagnetic ($J<0$) spin-spin coupling at magnetic fields that are considerably lower than those predicted by the standard HM model, in qualitative agreement with previously reported $sp$-hybridized calculations.\cite{Burkard_PRB99, Hu_PRA00} This behavior is consistent throughout the considered range of $d$ values and extends to biased double dots, as seen in figure \ref{Fig5}b. In contrast with the standard HM results, $J$ values calculated with our extended HM model show prominent ferromagnetic amplitudes, most notably in biased configurations, since we include triplet hybridization that lowers its energy and guarantees it remains the ground state over a wider range of magnetic fields.

\begin{figure}[t]
\epsfxsize=0.75\columnwidth
\vspace*{0.1 cm}
\centerline{\epsffile{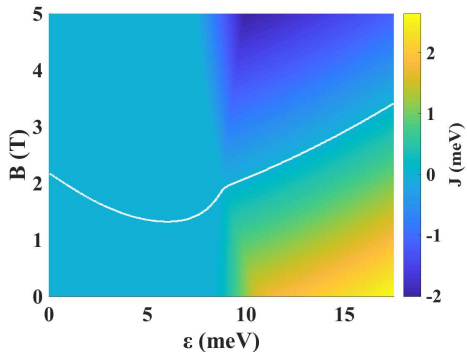}}
\vspace*{-0.15 cm}
\caption{(Color online) A color map of $J$ vs.~$B$ and $\varepsilon$ for $d=2.5$ and $\omega_0=5$ meV. The white line depicts singlet-triplet energy crossing ($J=0$). The $J$ values at the lower-bias regime are very small, as compared with the right side of the map, making it difficult to discern color variations in this regime.}
\label{Fig6}
\end{figure}

Figure \ref{Fig6} presents a color map of $J(B,\varepsilon)$, where the white line marking $J=0$, exhibits a non-monotonous dependence on bias. This behavior is associated with the different mechanisms that govern the $S-T$ energy crossing in the low- and high-bias regimes, as we now explain. In the low-bias regime, where both singlet and triplet states are predominantly in their separated (single occupancy) configurations, the $S-T$ crossing is governed by the competition of Coulomb exchange terms and direct (single-particle) tunneling.\cite{Hu_PRA00} Electrons in the triplet configuration tend to repel each other so that their Coulomb interaction is reduced, making the triplet the ground state when the Coulomb contribution dominates. As $B$ increases, orbital wavefunctions are squeezed and their overlap is reduced, making the long-range Coulomb couplings dominant. Within the low-bias regime, we find that the direct tunneling contribution becomes smaller as $\varepsilon$ increase (see table \ref{Tsingle} in appendix \ref{sec:app_Helements}), while Coulomb terms are independent of $\varepsilon$. As a result, lower $B$ is needed to establish Coulomb dominance (and thus negative $J$) as $\varepsilon$ is raised.

The origin of the increase in transition $B$ with bias at the high-bias regime is very different. Here, both singlet and triplet anticrossings have occurred and the two electrons predominantly occupy a single dot in both configurations. The ground orbital energy is increased with $B$ while the excited orbital energy is lowered. Together with a reduced triplet on-site Coulomb interaction (half in magnitude as compared with the singlet -- see rable \ref{TCoulomb} in appendix \ref{sec:app_Helements}), the triplet becomes the ground state at sufficiently high $B$. We find  different bias dependence of the singlet and triplet energies, arising from their single-particle $Q$ terms, where the excited orbital contribution decreases more slowly with $\varepsilon$ as compared with the ground orbital (see table \ref{Tsingle} in appendix \ref{sec:app_Helements}). As a result, when  $B$ is fixed at a value above the $S-T$ crossing and bias is swept up, the singlet energy will inevitably fall below the triplet energy. We stress again that the bias dependence at this regime originates from the details of the quartic potential and is therefore model-specific.
Finally, in the vicinity of the two anticrossings, the lowest-lying singlet and triplet states have sizable probabilities of both single- and double-occupancy orbital configurations, thus both aforementioned mechanisms impact the location of the $S-T$ energy crossing, resulting in a non-monotonous behavior in the bias range $5 \lessapprox \varepsilon \lessapprox 9$ mV, for the example depicted in figure \ref{Fig6}.

\begin{figure}[t]
\epsfxsize=0.75\columnwidth
\vspace*{0.1 cm}
\centerline{\epsffile{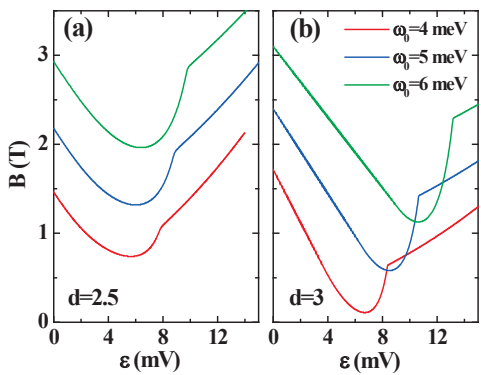}}
\vspace*{-0.2 cm}
\caption{(Color online) $B(\varepsilon)$ at which $J=0$ for several confinement energies with: (a) $d=2.5$, (b) $d=3$.}
\label{Fig7}
\end{figure}

The fact that the $S-T$ energy crossing takes place at lower magnetic fields in asymmetric double dot devices is practically useful, as it enables to turn off the exchange interaction by properly tuning the control fields to experimentally accessible values. The locations of this idle position for several device geometries are shown in figure \ref{Fig7}, where $B(\varepsilon)$ at which $J=0$ is plotted. At half interdot separation of $d=2.5$ (figure \ref{Fig7}a) we obtain the idle positions with minimum fields of $B=0.74$ T, 1.32 T, and 1.96 T at $\varepsilon =7$ mV, 6 mV, and 6.5 mV, respectively. For larger dot separation, $J=0$ is obtained at even lower magnetic fields, as depicted in figure \ref{Fig7}b, since the wavefunction overlap is lower to begin with, and lower $B$ is sufficient to compress the orbitals so that Coulomb interaction becomes dominant. While this trend is consistent for biased dots, we note that it is reversed at zero bias, where $J=0$ is reached at a higher field when $d$ is increased (see also figure \ref{Fig5}a). A closer inspection of $J(d)$ at zero bias reveals a non-monotonous approach to zero, a phenomenon that was also observed in Ref.~\onlinecite{Li_PRB10}, where it was hypothesized to be confinement-model-dependent. We conclude that, contrary to the biased case, the exact location of $J=0$ at zero bias is a result of a delicate balance between several contributions, and may well be the result of our use of quartic potential.

\section{Conclusion}
\label{Conclusion}

The exchange interaction between two electrons confined in a double QD is a key ingredient in all spin-based qubits, and its tunability through interdot bias provides a convenient control handle. At the same time, the exchange sensitivity to bias fluctuations, derived from the different charge distributions the singlet and triplet configurations, plays an important role in limiting the qubit coherence and reducing its gate fidelities.

In this work we developed an extended HM orbital model that improves the reliability and applicability range of the standard HM model. In particular, our model correctly accounts for the reduced sensitivity of $J$ to bias fluctuations in the high-bias regime, in excellent agreement with experimental values.\cite{Dial_PRL13} The analytical approach we have taken allowed us to study the dynamics of the exchange interaction throughout the validity range of our model, and identify the mechanisms behind it.

It would be interesting to use our extended HM model with a confinement potential that includes separate control over barrier height. This will allow us to model symmetric exchange control that was recently demonstrated\cite{Martins_PRL16,Reed_PRL16} and directly compare its performance with bias control, potentially unraveling useful information on the orbital landscape of double-dot devices. In addition, we expect that our extended HM approach can be applied to three-spin qubits in triple QDs,\cite{Laird_PRB10,Gaudreau_Nature11,Medford_Nature13,Medford_PRL13} possibly revealing subtle interplays between interdot tunnel couplings, intradot level splittings and spin correlation energies that directly impact exchange behavior in these devices. In this context, we mention a recent work that explored exchange coupling of a multielectron QD to a neighboring two-electron double dot.\cite{Martins_PRL17,Malinowski_arXiv17} Using a Hubbard model with an additional excited orbital, the authors were able to explain intricate non-monotonous exchange behavior (including negative $J$) for both spin-1/2 state at odd occupancies and spin-1 state at even occupancy of the multielectron dot.

\section*{ACKNOWLEDGEMENTS}

This work was supported by the National Science Foundation Grant no.~DMR 1207298.

\appendix

\section{Truncated Hilbert space under the Hund-Mulliken approximation}
\label{sec:app_truncate}

In this appendix we briefly discuss the applicability of our extended HM approach. A well-known validity test for exchange calculations is the Lieb-Mattis theorem, stating that at zero magnetic field the ground state of a two-particle system under a symmetric confinement potential must be in a singlet configuration.\cite{Mattis_Springer} It was noted that the Heitler-London model breaks down when $c=\sqrt{\pi/2} e^2/(\kappa a_B \hbar \omega_0) >2.8$ (corresponding to $\omega_0 =2.13$ meV for GaAs), predicting $J<0$ at $B=0$ for sufficiently large overlap between the two dots.\cite{Burkard_PRB99} Similarly, the standard HM breaks down, though at much smaller interdot distances. This problem intensifies for Si dots, where both the larger Coulomb energy (due to reduced screening) and the smaller kinetic energy (due to the larger effective mass) lead to critical $c$ values reached at smaller dot sizes.\cite{Li_PRB10}

The $p_-$ orbitals introduced in our extended HM model to allow for triplet hybridization in the high-bias regime. In addition, the $p_-$ orbitals induce anisotropy that allows for more spread in the electronic wavefunctions and a better account of the two-electron correlations, both contributing to lower the orbital energy.\cite{Hu_PRA00} The latter improvement should be particularly valuable as both Heitler-London and HM methods are known to underestimate electron correlations. While our extended HM approach should thus provide more reliable $J$ values in the smaller $d$ regime, it has been generally observed that the accuracy of CI calculations does not necessarily improve with the size of the basis states. An indication of the potential difficulties associated with state-basis truncation is given by our inability to orthogonalize the wavefunctions below $d \lessapprox 1.7$, largely independent of $\omega_0$. In this regime, neither the approximate equations, (\ref{gsol}), nor our numerical solver provide reliable values for the hybridization coefficients -- a problem that we believe is related to the absence of the $p_+$ orbital in our model.

\begin{figure}[t]
\epsfxsize=0.65\columnwidth
\vspace*{0.1 cm}
\centerline{\epsffile{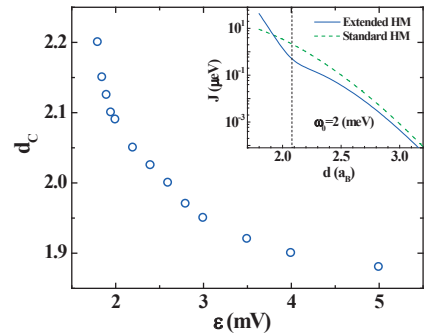}}
\vspace*{-0.1 cm}
\caption{(Color online) Critical $d$ values at which $\partial J/\partial d$ calculated with the extended HM model abruptly changes (but remains negative) against $\omega_0$. We judge that below $d_c$, calculated $J$ values are unreliable. The inset shows an example of $J(d)$ for $\omega_0=2$ meV, where both standard (dashed-green line) and extended (solid-blue line) HM results are shown. The vertical dashed line marks the reliability regime of the extended HM model.}
\label{Figappa}
\end{figure}

Steering away from this strong-coupling regime, which have not been implemented experimentally, to the best of our knowledge, we verify the applicability of our model in the rest of the $d$ and $\omega_0$ range, by examining the values for which $\partial J/\partial d$ decreases monotonically with $d$ at $B=0$ and $\varepsilon =0$. We find that for $\omega_0 \approx 1.6$ meV (corresponding to $a_B \approx 270 \AA$ and $c \approx 3.25$), $\partial J/\partial d$ becomes positive at $d\approx 2.4$, marking a breakdown of our model. This critical $d$ value is increased for yet smaller $\omega_0$ values (or larger dots). In this regime the standard HM model performs better then our extended HM model, breaking down at $d=1.65$ for $\omega_0=1.6$ meV. On the other hand, for $\omega_0 >1.6$ meV, the extended HM model does not generate positive $\partial J/\partial d$ for any $d$ value. Instead, as $d$ is reduced, $\partial J/\partial d$ presents an abrupt increase in magnitude at some critical value $d_c$, that is reduced with increased $\omega_0$, as depicted in figure \ref{Figappa}. While not in violation with the Lieb-Mattis theorem, we believe that this behavior is an artifact of our truncated space, and conclude that $J$ values in this regime are less reliable. All results presented in the main text are for QDs with confinement energies above $\omega_0=4$ meV and half separation above $d=2.5$, well above the validity limits indicated by figure \ref{Figappa}.

Another limitation of our model emerges at very large magnetic fields, where $J$ drops exponentially with $B$ (see, e.g., figure \ref{Fig5}).
A numerical calculation that studied corrections to the Heisenberg interaction revealed that at high magnetic fields, the ground state alternates between singlet and triplet states, so that additional zero crossings can occur at larger $B$ fields than those shown in figures \ref{Fig5}-\ref{Fig7} in the main text.\cite{Scarola_PRA05}

By far, our crudest approximation is to exclude the relatively close $p_+$ orbital in our calculations. The energy splitting between the $p_\pm$ orbitals is $\approx 0.35 \hbar \omega_0$ at $B=1$ T, and while $J$ can get as high as a $1-2$ meV at the large-bias regime, none of the direct tunneling or Coulomb terms exceeds $\approx 10 \mu$eV for $d \geq 2.5$. The truncation of the $p_+$ orbital is not a fundamental limitation of our approach and it is possible to include it, albeit at the price of additional hybridization coefficients and an increase in Hilbert space dimension from 12 to 20 (11 Singlets and 9 triplets, provided that we keep two-electron states with only one $p$ orbital). While not fully justified, our truncated Hilbert space results in exchange values that pass all validity tests within the parameter range used, and match experimental values surprisingly well.

\section{Orbital Hamiltonian matrix elements}
\label{sec:app_Helements}
	
In this appendix we provide details of the calculation of the matrix elements of our orbital Hamiltonian. It is convenient to evaluate the single-particle portion of the Hamiltonian, Eq.~(\ref{H}), by referencing it to the Hamiltonian of two-dimensional harmonic wells, centered at $\pm d$,\cite{Burkard_PRB99} such that
\begin{eqnarray}
H_{\rm orb}\!&\!=\!&\! h_{-}({\bf r}_1)+h_{+}({\bf r}_2)+Q_-({\bf r}_1)+Q_+({\bf r}_2)+H_C \nonumber \\
h_{\pm}({\bf r}) \!&\!=\!&\! \frac{1}{2} \left( {\bf p}-\frac{e}{c} {\bf A} ({\bf r}) \right)^2
\!\!+\! \frac{1}{2} \left[ (x\mp d)^2+y^2 \right] \!+\! \frac{x \varepsilon}{d} \\
Q_\pm ({\bf r}) \!&\!=\!&\! \frac{1}{2} \left[ \frac{x^4}{4d^2}-\frac{3x^2}{2}-\frac{3d^2}{4} \pm 2xd \right], \nonumber
\end{eqnarray}
where coordinates are measured in $a_B$ units, generalized momenta in $\hbar a_B^{-1}$ units, and energies in $\hbar \omega_0$ units.

The calculation of the system's eigenenergies is carried out in four steps: (i) closed-form expressions for the single-particle (kinetic) and two-particle (Coulomb) bare matrix elements, are calculated within the ground and excited Fock-Darwin orbitals, Eqs.~(\ref{FD}), (ii) The orthogonalized matrix elements are constructed out of combinations of the bare terms, according to Eqs,~(\ref{Phi}), (iii) the orthogonalized terms are combined to find the final Hamiltonian
matrix elements for the 12 singlet and triplet states listed in table \ref{Tstates}, (iv) the resulting matrix is diagonalized to find the orbital eigenenergies. Below we provide details pertaining to these calculational steps.

\subsection{Single-particle bare matrix elements}

We define single-particle energy and tunneling terms in a biased quartic potential as:
\begin{eqnarray}
\epsilon_\pm^g&=& \bra{\varphi^g_\pm} h_\pm+Q_\pm \ket{\varphi^g_\pm} \notag \\
\epsilon_\pm^e&=& \bra{\varphi^e_\pm} h_\pm+Q_\pm \ket{\varphi^e_\pm} \notag \\
\epsilon_\pm^{ge}&=& \bra{\varphi^g_\pm} h_\pm+Q_\pm \ket{\varphi^e_\pm},
\end{eqnarray}
and
\begin{eqnarray}
t^g &=& \bra{\varphi^g_\pm} h_\pm+Q_\pm \ket{\varphi^g_\mp} \notag \\
t^e &=& \bra{\varphi^e_\pm} h_\pm+Q_\pm \ket{\varphi^e_\mp} \notag \\
t_\pm^{ge}&=& \bra{\varphi^g_\pm} h_\pm+Q_\pm \ket{\varphi^e_\mp},
\label{t}
\end{eqnarray}
where $h_\pm$ and $Q_\pm$ are the (biased) harmonic-well Hamiltonians and quartic potential correction terms, respectively. Note that for all three cases (ground-, excited- or mixed-orbitals) we have $\bra{\varphi_\pm} h_\mp+Q_\mp \ket{\varphi_\pm}=\bra{\varphi_\mp} h_\mp+Q_\mp \ket{\varphi_\mp}$ and $\bra{\varphi_\pm} h_\mp+Q_\mp \ket{\varphi_\mp}=\bra{\varphi_\mp} h_\mp+Q_\mp \ket{\varphi_\pm}$ (although the $h$ and $Q$ matrix elements are not equal independently, their combinations are). In addition, in Eqs.~(\ref{t}) we used the fact that $t^g_+=t^g_-\equiv t^g$ and $t^e_+=t^e_- \equiv t^e$. The explicit expressions for the $h$ and $Q$ matrix elements are listed in table \ref{Tsingle}.
\begin{longtable*}{l}
\caption{Bare single-particle matrix elements in $\hbar \omega_0$ units. The overlap integrals are given in Eqs.~(\ref{S}).}
\label{Tsingle} \\
\hline \hline \\[1pt]
$h$ terms \\[1pt] \hline \\[1pt]
$\bra{\varphi_{\pm}} h_{\pm} \ket{\varphi_{\pm}}=b - \frac{\varepsilon^2}{8d^2} \pm \frac{\varepsilon}{2}$ \\
$\bra{\varphi_{\pm}} h_{\pm} \ket{\varphi_{\mp}}=S\left(b - \frac{\varepsilon^2}{8d^2} \pm \frac{\varepsilon}{2}\right)$ \\
$\bra{\varphi_{\pm}^e} h_{\pm} \ket{\varphi_{\pm}^e}=2b - \sqrt{b^2-1} - \frac{\varepsilon^2}{8d^2} \pm \frac{\varepsilon}{2}$ \\
$\bra{\varphi_{\pm}^e} h_{\pm} \ket{\varphi_{\mp}^e}=S_{ee}\left(2b-\sqrt{b^2-1}-\frac{\varepsilon^2}{8d^2} \pm \frac{\varepsilon}{2}\right)$ \\
$\bra{\varphi_{\pm}} h_{\pm} \ket{\varphi_{\mp}^e}=S_{\pm}\left(b - \frac{\varepsilon^2}{8d^2} \pm \frac{\varepsilon}{2}\right)$\\
$\bra{\varphi_{\pm}^e} h_{\pm} \ket{\varphi_{\mp}} = S_{\mp}\left(2b-\sqrt{b^2-1}-\frac{\varepsilon^2}{8d^2} \pm \frac{\varepsilon}{2}\right)$ \\
$\bra{\varphi_{\pm}} h_{\pm} \ket{\varphi_{\pm}^e} = 0$ \\
$\bra{\varphi_{\pm}} h_{\mp} \ket{\varphi_{\pm}^e} = \mp \frac{d}{\sqrt{b}}$ \\[1pt]
\hline \\[1pt]
$Q$ terms \\[1pt] \hline \\[1pt]
$\bra{\varphi_{\pm}} Q_{\pm} \ket{\varphi_{\pm}} = \frac{1}{32b^2d^4} \left[ 3d^2 + 3b \varepsilon (\varepsilon \mp 4d^2)+\frac{b^2 \varepsilon^3}{4d^2} (\varepsilon \mp 8d^2)   \right]$ \\
$\bra{\varphi_{\pm}} Q_{\pm} \ket{\varphi_{\mp}} = \frac{S}{32b^2d^4} \left[ 3d^2 + 3b(\varepsilon^2 - 4d^4)- \frac{b^2}{4d^2}(2d^2 \pm \varepsilon)^3(6d^2 \mp \varepsilon) \right]$ \\
$\bra{\varphi_{\pm}^e} Q_{\pm} \ket{\varphi_{\pm}^e} = \frac{1}{32b^2d^4} \left[ 9d^2+6b \varepsilon (\varepsilon \mp 4d^2)+\frac{b^2}{4d^2} \varepsilon^3(\varepsilon \mp 8d^2)  \right]$ \\
$\bra{\varphi_{\pm}^e} Q_{\pm} \ket{\varphi_{\mp}^e} = \frac{1}{32b^2 d^4} \left\{ S_{ee}\left[3d^2-\frac{b^2}{4d^2}(2d^2\pm \varepsilon)^3 (6d^2 \mp \varepsilon) + 3b(\varepsilon^2 -4d^4) \right] +3S(2d^2+b\varepsilon^2 - 2d^4 ) \right\}$ \\
$\bra{\varphi_{\pm}} Q_{\pm} \ket{\varphi_{\mp}^e} = \frac{S_{\pm}}{32b^2 d^4} \left[3d^2-\frac{b^2}{4d^2}(2d^2\pm\varepsilon)^3(6d^2 \mp \varepsilon) + 4b(\varepsilon^2 -4d^4) \right]  - \frac{S}{16d^3b^{3/2}} \left[3\varepsilon + \frac{b}{2d^2}(\varepsilon \mp 4d^2)(\varepsilon \pm 2d^2)^2 \right]$ \\
$\bra{\varphi_{\pm}^e} Q_{\pm} \ket{\varphi_{\mp}} = \frac{S_{\mp}}{32b^2 d^4} \left[3d^2-\frac{b^2}{4d^2}(2d^2\pm\varepsilon)^3(6d^2 \mp \varepsilon) + 4b(\varepsilon^2 -4d^4) \right]  - \frac{S}{16d^3b^{3/2}} \left[3\varepsilon + \frac{b}{2d^2}(\varepsilon \mp 4d^2)(\varepsilon \pm 2d^2)^2 \right]$ \\
$\bra{\varphi_{\pm}} Q_{\pm} \ket{\varphi_{\pm}^e} = -\frac{1}{16d^3 b^{3/2}} \left[3(\varepsilon \mp 2d^2)+ \frac{b}{2d^2}\varepsilon^2(\varepsilon \mp 6d^2) \right]$ \\
$\bra{\varphi_{\pm}} Q_{\mp} \ket{\varphi_{\pm}^e} = -\frac{1}{16d^3 b^{3/2}} \left[3(\varepsilon \pm 2d^2)+ \frac{b}{2d^2}\varepsilon^2(\varepsilon \pm 6d^2)\mp 16bd^4 \right]$ \\[4pt]
\hline \hline
\end{longtable*}

\subsection{Coulomb Terms}

We denote a generic bare (non-orthogonalized) Coulomb term as $c^{ijkl}$, where $i,j,k,l \in \left\{ g,e \right\}$ and $c \in \left\{ u,v_{\rm d},v_{\rm x},w,x \right\}$. The five types of Coulomb couplings correspond to the various double dot occupancies (in either ground or excited orbitals) and are given by:
\begin{eqnarray}
u \!&\!=\!&\! \bra{\varphi_+ \varphi_+} H_C \ket{\varphi_+ \varphi_+} \notag \\
v_{\rm d} \!&\!=\!&\! \bra{\varphi_+ \varphi_-} H_C \ket{\varphi_+ \varphi_-} \notag \\
v_{\rm x} \!&\!=\!&\! \bra{\varphi_+ \varphi_-} H_C \ket{\varphi_- \varphi_+} \notag \\
w \!&\!=\!&\! \bra{\varphi_+ \varphi_+} H_C \ket{\varphi_+ \varphi_-} \notag \\
x \!&\!=\!&\! \bra{\varphi_+ \varphi_+} H_C \ket{\varphi_- \varphi_-},
\label{udvwx}
\end{eqnarray}
where $\varphi_\pm \in \left\{ \varphi_\pm^g, \varphi_\pm^e \right\}$. Symmetry properties of the Coulomb interaction, $\bra{\varphi_i \varphi_j} H_C\ket{\varphi_k \varphi_l} = \bra{\varphi_j \varphi_i} H_C\ket{\varphi_l \varphi_k} =\bra{\varphi_k \varphi_l} H_C\ket{\varphi_i \varphi_i}$, determine the number of distinct matrix elements in each of the five groups. In addition, it can be verified that the simultaneous swapping of all four orbitals, $+ \longleftrightarrow -$, in Eqs.~(\ref{udvwx}) leaves the term unchanged for an even number of excited orbitals and adds a sign for an odd number of excited orbitals. Although our Hilbert space is truncated to include up to one excited orbital in each dot, the orthogonalized matrix elements include (small) contributions from all 16 bare orbital combinations. Taking these symmetry considerations into account we find that there are seven distinct terms of each of the $u$, $v_{\rm d}$, $v_{\rm x}$, and $x$ types, and sixteen distinct terms of the $w$ type. The explicit closed-form expressions for these bare Coulomb terms are listed in table \ref{TCoulomb}.

\begin{longtable*}{l}
\caption{Bare Coulomb terms in $\hbar \omega_0 c$ units (only distinct terms are listed). The overlap integrals are given in Eqs.~(\ref{S}). In these expressions, $I_n^d\equiv I_n(bd^2)$, $I_n^x \equiv I_n \left[ d^2(b-1/b)\right]$, and $I^w_n \equiv I_n (d^2/4b ) $ denote the $n$th order modified Bessel functions with the indicated arguments.}
\label{TCoulomb} \\
\hline \hline \\[1pt]
$u$ terms \\[1pt] \hline \\[1pt]
$u^{gggg}=\sqrt{b}$ \\ $u^{eggg}=0$ \\ $u^{egeg}=\frac{3}{4} \sqrt{b}$ \\
$u^{egge}=\frac{1}{4} \sqrt{b}$ \\ $u^{eegg}=0$ \\ $u^{eeeg}=0$ \\
$u^{eeee}=\frac{11}{16} \sqrt{b}$\\[1pt] \hline \\[1pt]
$v_{\rm d}$ terms \\[1pt] \hline \\[1pt]
$v_{\rm d}^{gggg}=\sqrt{b} e^{-bd^2} I^d_0$ \\
$v_{\rm d}^{eggg}=-\frac{bd}{2} e^{-bd^2}\left[I^d_0-I^d_1\right]$ \\
$v_{\rm d}^{egeg}=\frac{\sqrt{b}}{4} e^{-bd^2} \left[3I^d_0 +2bd^2(I^d_0-I^d_1) \right]$ \\
$v_{\rm d}^{egge}=\frac{\sqrt{b}}{4} e^{-bd^2} \left[I^d_0- 2bd^2(I^d_0-I^d_1)\right]$ \\
$v_{\rm d}^{eegg}=\frac{\sqrt{b}}{4} e^{-bd^2} \left[I^d_1-2bd^2 (I^d_0-I^d_1)\right]$\\
$v_{\rm d}^{eeeg}=\frac{bd}{2} e^{-bd^2}\left[ bd^2(I^d_0-I^d_1)+\frac{1}{4}(I^d_0-3I^d_1)\right]$ \\
$v_{\rm d}^{eeee}=\frac{\sqrt{b}}{4} e^{-bd^2} \left[ \left(\frac{11}{4}-bd^2 \right)I^d_0+2bd^2(1+bd^2)(I^d_0-I^d_1) \right]$ \\[1pt]
\hline \\[1pt]
$v_{\rm x}$ terms \\[1pt] \hline \\[1pt]
$v_{\rm x}^{gggg}=\sqrt{b} e^{-bd^2} S I^x_0$ \\
$v_{\rm x}^{eggg}=-  \frac{\sqrt{b}}{2} e^{-bd^2} \left[ bS (I^x_0+I^x_1)+S_+ (I^x_0 -I^x_1) \right]$  \\
$v_{\rm x}^{egeg}=\frac{\sqrt{b}}{2} e^{-bd^2} \left[ S_{ee}(I^x_0-I^x_1)+ S(1-bd^2)(I_0^x+I_1^x)-\frac{1}{2}S I^x_0 \right]$ \\
$v_{\rm x}^{egge}=\frac{b^{3/2}}{2d^2} e^{-bd^2} \left[bd^2 S(I^x_0+I^x_1)+(S-S_{ee}) (I^x_0-I^x_1)+\frac{1}{2}SI^x_0 \right]$ \\
$v_{\rm x}^{eegg}=-\frac{b^{3/2}}{2d^2} e^{-bd^2} \left[bd^2 S(I^x_0+I^x_1)+(S-S_{ee}) (I^x_0-I^x_1)-\frac{1}{2}SI^x_1 \right]$ \\
$v_{\rm x}^{eeeg}=\frac{\sqrt{b}}{2} e^{-bd^2} \left\{ \left[ (b+\sqrt{b^2-1})S_{ee}-\frac{b}{4} S-\frac{1}{2}S_+ \right](I^x_0-I^x_1)+\left[ \frac{b}{2}S+\frac{1}{4}S_+-b^2d^2S \right] (I^x_0+I^x_1)\right\}$ \\
$v_{\rm x}^{eeee}=\frac{\sqrt{b}}{4} e^{-bd^2} \left\{ \left[ \left( \frac{11}{8} +bd^2\right) S- 2d^2\left( 2b +\frac{1}{b}+2\sqrt{b^2-1} \right)S_{ee}\right] (I^x_0-I_1^x)+\left[\frac{3}{8}S+S_{ee}-d^2\left(b+\frac{1}{b}-2b^2d^2\right)S\right] (I_0^x+I_1^x) \right\}$ \\[1pt]
\hline \\[1pt]
$w$ terms \\[1pt]
\hline \\[1pt]
$w^{gggg} = \sqrt{b} e^{\frac{-d^2}{4b}} S I^w_0$ \\
$w^{eggg} =  -\frac{\sqrt{b}}{4}e^{-\frac{d^2}{4b}} S \left(I^w_0 - I^w_1 \right)$ \\
$w^{gegg} = \frac{\sqrt{b}}{4}e^{-\frac{d^2}{4b}} S \left(3I^w_0 + I^w_1 \right)$ \\
$w^{ggeg} =  -\frac{\sqrt{b}}{4}e^{-\frac{d^2}{4b}} \left(2bS-S_+\right)\left(I^w_0 - I^w_1 \right)$ \\
$w^{ggge} = \frac{\sqrt{b}}{4}e^{-\frac{d^2}{4b}} \left[ 2bS \left(I^w_0 - I^w_1 \right)+ S_+\left(3I^w_0 + I^w_1 \right) \right]$ \\
$w^{egeg} = \frac{1}{8 \sqrt{b}} e^{-\frac{d^2}{4b}} S \left(d^2(I^w_0-I^w_1)+6bI^w_0 \right)$ \\
$w^{gege} = \frac{1}{8 \sqrt{b}} e^{-\frac{d^2}{4b}} \left[ d^2(5I^w_0+3I^w_1)S-4bd^2 (3I^w_0 + I^w_1)S_+ + 6bI^w_0S\right]$ \\
$w^{egge} = \frac{1}{8 \sqrt{b}} e^{-\frac{d^2}{4b}} \left[ d^2\left(S-4bS_+\right) (I^w_0 - I^w_1) + 2bI^w_0 S\right]$ \\
$w^{geeg} = \frac{1}{8 \sqrt{b}} e^{-\frac{d^2}{4b}} S \left[ d^2(I^w_0 - I^w_1)+ 2bI^w_0 \right]$ \\
$w^{eegg} =  \frac{1}{8d} e^{-\frac{d^2}{4b}} S_+ (b+\sqrt{b^2-1})  \left[ d^2(I^w_0 - I^w_1)+ 2bI^w_0  \right]$ \\
$w^{ggee} = \frac{1}{8 \sqrt{b}} e^{-\frac{d^2}{4b}} \left\{ \left[2d^2S+b(S-S_{ee})\right] (I^w_1 - I^w_0) + 2\frac{b^2}{d^2}(S-S_{ee}) I^w_1  \right\}$ \\
$w^{eeeg} = \frac{1}{16 \sqrt{b}} e^{-\frac{d^2}{4b}} S_+ \left[ d^2(I^w_1-I^w_0)-b(11I^w_0 + 3I^w_1)  \right]$ \\
$w^{eege} =  \frac{1}{16 \sqrt{b}} e^{-\frac{d^2}{4b}} \left\{ d^2\left[ (8b^2-1)S_+ -4bS\right](I^w_0 - I^w_1 ) -5bS_+(I^w_0 + I^w_1) + 4b\left[(4b^2+1)S_+-2bS\right] I^w_1  \right\}$ \\
$w^{egee} = \frac{1}{16 \sqrt{b}} e^{-\frac{d^2}{4b}} \left\{\left[ (1+2b^2)(4b+d^2)S+2bd^2(bS-S_+)-4b^2S_+\right](I^w_0 - I^w_1) + \left[7bS-2b^2(2bS-S_+)\right](I^w_0 + I^w_1) \right\}$ \\
$w^{geee} = \frac{1}{16 \sqrt{b}} e^{-\frac{d^2}{4b}} \left\{ \left[ (d^2+2b)S_+ +2bd^2S \right] (I^w_0 - I^w_1) + (3bS_+ -2b^2S)(I^w_0 + I^w_1) \right\}$ \\
$w^{eeee} =  \frac{1}{32 \sqrt{b}} e^{-\frac{d^2}{4b}} \left\{ \left[2(d^2-10b)S_{ee}-d^2\left(1+\frac{d^2}{b}\right)S \right](I^w_0 - I^w_1) - \left[14b S_{ee}-3d^2S \right] (I_0^w+I_1^w) +2b \left(3-28bd^2 \right)S I_0^w \right\}$ \\[1pt]
\hline \\[1pt]
$x$ terms \\[1pt]
\hline \\[1pt]
$x^{gggg} =  \sqrt{b} S^2$ \\
$x^{eggg} = -\sqrt{b} SS_{+}$ \\
$x^{egeg} = \sqrt{b} S\left(S_{ee}+\frac{1}{4}S\right)$ \\
$x^{egge} = \sqrt{b} S_+ \left[\frac{b^2}{2d^2}S-\left(1+\frac{b}{4d^2}\right) S_+ \right]$ \\
$x^{eegg} = \sqrt{b} S_+^2$ \\
$x^{eeeg} = -\sqrt{b} S_{ee}S_+$ \\
$x^{eeee} = \sqrt{b} \left( S_{ee}^2-\frac{5}{16}S^2\right)$ \\[1pt]
\hline \hline
\end{longtable*}

\subsection{Final orbital matrix elements}
The terms listed in Tables \ref{Tsingle} and \ref{TCoulomb} and the hybridization coefficients, Eqs.~(\ref{gsol}), are used to construct the orthogonalized kinetic and Coulomb matrix elements. We denote orthogonalized elements with capital letters, i.e., $E$ and $T$ for energy and tunneling terms, and $U, V_{\rm d}, V_{\rm x}, W$, and $X$ for Coulomb terms. These orthogonalized elements are then combined to provide the final matrix elements of the orbital Hamiltonian, in the basis of the 12 singlet and triplet states listed in table \ref{Tstates}. Since singlets are symmetric and triplets are antisymmetric in their orbital degrees of freedom, they are not coupled by our symmetric Hamiltonian, resulting in a total of 28 singlet and 15 triplet matrix elements, listed in table \ref{THterms}. Explicit expressions of elements involving only ground orbitals can be found in ref.~\onlinecite{Burkard_PRB99} (for non-biased configuration) and in ref.~\onlinecite{Ramon_arXiv09} (including bias).

\begin{longtable*}{l|l}
\caption{Orbital Hamiltonian matrix elements. In these terms $V_\pm \equiv V_{\rm d} \pm V_{\rm x}$.}
\label{THterms} \\
\hline \hline
\multicolumn{2}{c}{} \\[1pt]
\multicolumn{2}{c}{Diagonal Terms} \\[1pt] \hline \\[1pt]
Singlets & Triplets \\[1pt] \hline \\[1pt]
$E_{S(0,2)}=2E_+^g + U^{gggg}$ & --- \\
$E_{S(2,0)}=2E_-^g + U^{gggg}$ & --- \\
$E_{S(1,1)}=E_+^g + E_-^g + V_+^{gggg} $ & $E_{T(1,1)}=E_+^g + E_-^g + V_-^{gggg}$ \\
$E_{S_{e}(0,2)}=E_+^g+E_+^e + U^{egeg}+U^{egge}$ & $E_{T_e(0,2)}=E_+^g+E_+^e + U^{egeg}-U^{egge}$\\
$E_{S_{e}(2,0)}=E_-^g + E_-^e + U^{egeg}+U^{egge}$ & $E_{T_e(2,0)}=E_-^g + E_-^e + U^{egeg}-U^{egge}$\\
$E_{S_{ge}(1,1)}=E_-^g + E_+^e +V_{\rm d}^{egeg}+V_{\rm x}^{egge}$ & $E_{T_{ge}(1,1)}=E_-^g + E_+^e +V_{\rm d}^{egeg}-V_{\rm x}^{egge}$ \\
$E_{S_{eg}(1,1)}=E_+^g + E_-^e +V_{\rm d}^{egeg}+V_{\rm x}^{egge}$ & $E_{T_{eg}(1,1)}=E_+^g + E_-^e +V_{\rm d}^{egeg}-V_{\rm x}^{egge}$ \\[1pt] \hline
\multicolumn{2}{c}{} \\[1pt]
\multicolumn{2}{c}{Off-diagonal Terms} \\[1pt] \hline \\[1pt]
Singlets & Triplets \\[1pt] \hline \\[1pt]
$\bra{S(0,2)} H \ket{S(2,0)} = X^{gggg}$ & --- \\
$\bra{S(0,2)} H \ket{S(1,1)} = \sqrt{2}(T^g+W^{gggg})$  & --- \\
$\bra{S(0,2)} H \ket{S_e(0,2)} = \sqrt{2}(E_+^{ge} + U^{eggg})$  & --- \\
$\bra{S(0,2)} H \ket{S_e(2,0)} = \sqrt{2}X^{ggge}$ & ---  \\
$\bra{S(0,2)} H \ket{S_{ge}(1,1)} = \sqrt{2}W^{ggeg}$ & ---  \\
$\bra{S(0,2)} H \ket{S_{eg}(1,1)} = \sqrt{2}(T_+^{ge}+W^{ggge})$ & ---  \\
$\bra{S(2,0)} H \ket{S(1,1)} = \sqrt{2}(T^g+W^{gggg})$  & --- \\
$\bra{S(2,0)} H \ket{S_e(0,2)} = \sqrt{2}X^{eggg}$  & --- \\
$\bra{S(2,0)} H \ket{S_e(2,0)} = \sqrt{2}(E_-^{ge}+U^{eggg})$ & --- \\
$\bra{S(2,0)} H \ket{S_{ge}(1,1)} = \sqrt{2}(T_-^{ge} - W^{ggge})$ & ---  \\
$\bra{S(2,0)} H \ket{S_{eg}(1,1)} = -\sqrt{2}W^{ggeg}$  & --- \\
$\bra{S(1,1)} H \ket{S_e(0,2)} = T_-^{ge}+W^{gegg}+W^{eggg}$ &  $\bra{T(1,1)} H \ket{T_e(0,2)} = T_-^{ge}+W^{gegg}-W^{eggg}$ \\
$\bra{S(1,1)} H \ket{S_e(2,0)} = T_+^{ge}-W^{gegg}-W^{eggg}$ &  $\bra{T(1,1)} H \ket{T_e(2,0)}  = T_+^{ge}+W^{gegg}-W^{eggg}$ \\
$\bra{S(1,1)} H \ket{S_{ge}(1,1)} = E_+^{ge}+V_+^{eggg}$ &  $\bra{T(1,1)} H \ket{T_{ge}(1,1)} = E_+^{ge}+V_-^{eggg}$ \\
$\bra{S(1,1)} H \ket{S_{eg}(1,1)} = E_-^{ge}+V_+^{eggg}$ & $\bra{T(1,1)} H
\ket{T_{eg}(1,1)}  = -(E_-^{ge}+V_-^{eggg})$ \\
$\bra{S_e(0,2)} H \ket{S_e(2,0)} = X^{egeg}+X^{egge}$ & $\bra{T_e(2,0)} H \ket{T_e(2,0)} = X^{egeg}-X^{egge}$\\
$\bra{S_e(0,2)} H \ket{S_{ge}(1,1)} = T^{g} + W^{egeg}+W^{geeg}$ &  $\bra{T_e(0,2)} H \ket{T_{ge}(1,1)} = T^g+W^{egeg}-W^{geeg}$ \\
$\bra{S_e(0,2)} H \ket{S_{eg}(1,1)} = T^e+W^{gege}+W^{egge}$ &  $\bra{T_e(0,2)} H \ket{T_{eg}(1,1)}=T^e+W^{egge}-W^{gege}$ \\
$\bra{S_e(2,0)} H \ket{S_{ge}(1,1)} = T^e-(W^{egge}+W^{gege})$ & $\bra{T_e(2,0)} H \ket{T_{eg}(1,1)} = T^e-(W^{gege}-W^{egge})$ \\
$\bra{S_e(2,0)} H \ket{S_{eg}(1,1)} = T^g-(W^{geeg}+W^{egeg}$ & $\bra{T_e(2,0)} H \ket{T_{eg}(1,1)} = -T^g-(W^{geed}-W^{egeg})$ \\
$\bra{S_{ge}(1,1)} H \ket{S_{eg}(1,1)} = V_{\rm d}^{egge} +V_{\rm x}^{egeg}$ & $\bra{T_{ge}(1,1)} H \ket{T_{eg}(1,1)} = V_{\rm d}^{egge}-V_{\rm x}^{egeg}$ \\[2pt]
\hline \hline
\end{longtable*}

\section{Local analytical approximations}
\label{sec:app_approx}

\begin{figure}[t]
\epsfxsize=0.65\columnwidth
\vspace*{0.3 cm}
\centerline{\epsffile{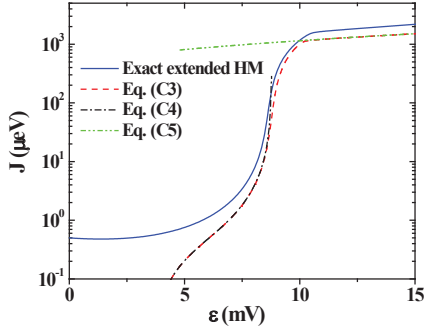}}
\vspace*{-0.3 cm}
\caption{(Color online) Exact solution for $J(\varepsilon)$ (solid blue line) and several approximations  in the vicinity of the singlet and triplet anticrossing, discussed in the text. Parameters used are: $B=0.1$ T, $\omega_0=5$ meV and $d=2.5$.}
\label{Figappc}
\end{figure}

The extended HM model used to generate the results presented in the main text includes 12 two-particle basis states given in table \ref{Tstates}. In the vicinity of the singlet and triplet anticrossings, the higher states become sufficiently removed from the lowest lying levels and one can obtain approximate exchange values by considering the reduced $2 \times 2$ singlet and triplet subspaces comprising of $S(1,1)-S(2,0)$ and $T(1,1)-T_e(2,0)$, respectively.  The singlet and triplet Hamiltonians are given by
\begin{equation}
H_S = \left(
\begin{array}{cc}
E_{S(2,0)} & \sqrt{2}(T^g+W) \\
\sqrt{2}(T^g+W) & E_{S(1,1)}
\end{array}
\right)
\end{equation}
and
\begin{equation}
H_T\!=\! \left(
\begin{array}{cc}
E_{T_e(2,0)} & T_+^{ge}+\Delta W \\
T_+^{ge}+\Delta W & E_{T(1,1)}
\end{array} \right),
\end{equation}
where the two-particle energies are given in table \ref{THterms}, the direct tunneling and Coulomb terms are given in tables \ref{Tsingle} and \ref{TCoulomb}, respectively, and we have defined $\Delta W \equiv W^{gegg}-W^{eggg}$ and omitted superscripts from ground-orbital-only Coulomb matrix elements for brevity. Diagonalizing these Hamiltonians, we find the approximate $J$ as
\begin{eqnarray}
J\!&= &\! E_{-}^T-E_-^S=\frac{1}{2} \left[ E_{T_e(2,0)}\!+\!E_{T(1,1)}\!-\!E_{S(2,0)}\!-\!E_{S(1,1)} + \right. \notag \\ && \left. \sqrt{\left(E_{S(2,0)}\!-E_{S(1,1)}\right)^2\!+8 (T^g\!+W)^2} - \right. \notag \\ && \left. \sqrt{\left(E_{T_e(2,0)}\!- E_{T(1,1)}\right)^2\! +4(T^{ge}\!+\Delta W)^2} \right],
\label{Japp}
\end{eqnarray}
depicted by the red-dashed line in figure \ref{Figappc}. Keeping only leading contributions in interdot overlap (appropriate for the considered weak-coupling regime), we find the asymptotes of Eq.~(\ref{Japp}) at biases below the singlet anticrossing ($\varepsilon_{\rm SC}$) and above the triplet anticrossing ($\varepsilon_{\rm TC}$) as:
\begin{equation}
J \approx \left\{ \begin{array}{ll}\!\!\! -2V_{\rm x}+\frac{2(t^g+w)^2}{\epsilon_-^g-\epsilon_+^g+u-v_+} - \frac{(t_+^{ge}+\Delta w)^2}{\epsilon_-^e-\epsilon_+^g+u^{egeg}-v_+} & , \varepsilon < \varepsilon_{\rm SC}\\
\epsilon_-^e-\epsilon_-^g-u+u^{egeg} & , \varepsilon > \varepsilon_{\rm TC}. \end{array} \right.
\label{Jappa}
\end{equation}
We note that in the lower bias regime, the first two terms in Eq.~(\ref{Jappa}) match the extended Hubbard limit given in Ref.~\onlinecite{Burkard_PRB99}, with interdot Coulomb correlations (first term) and tunneling and on-site Coulomb repulsion renormalized by long-range Coulomb couplings (second term). The third term in the lower bias regime of  Eq.~(\ref{Jappa}) is a new contribution due to the triplet hybridization, whose magnitude is comparable to those of the other terms, for the considered parameter range. At lower bias, where $J$ becomes very small, long-range contributions from higher states, absent in this reduced Hamiltonian picture, become important and the approximation breaks down, as seen by the dotted-green line in figure \ref{Figappc}.

At the high bias regime we find that the asymptotic behavior of  Eq.~(\ref{Japp}) is accurately captured by the energy difference between the lowest lying singlet and triplet configurations, since both are predominantly doubly-occupied. Using the explicit expressions for single-particle energies and bare Coulomb terms, given in tables \ref{Tsingle} and \ref{TCoulomb}, respectively, the resulting approximate $J$ at this high-bias limit is given by
\begin{eqnarray}
J_{\rm a}^{\rm hb} \!&\! \approx & \hbar \omega_0 \left[b- \! \sqrt{b^2-1}+\frac{3}{16b^2 d^2}-\frac{c\sqrt{b}}{2} \right] + \notag \\ && \frac{3}{8bd^2} \left(\varepsilon+\frac{\varepsilon^2}{4d^2 \hbar \omega_0} \right).
\label{Japphb}
\end{eqnarray}
While this result suggests that there is no sweet spot, at which $\partial J /\partial \varepsilon = 0$, we note that the bias dependence in Eq.~(\ref{Japphb}) emerges from the quartic-potential-related $Q$ terms and is therefore specific to our confinement model.

\end{document}